\documentclass[final,5p,times,twocolumn]{elsarticle}

\usepackage{amssymb}
\usepackage{amsmath}
\usepackage{algorithm}
\usepackage{algpseudocode}
\usepackage{natbib}
\usepackage{graphicx}
\usepackage{subcaption}
\usepackage{multirow}
\usepackage{bm}
\usepackage{booktabs}
\journal{Expert Systems with Applications}

\begin{document}

\begin{frontmatter}




\title{NASPrecision: Neural Architecture Search-Driven Multi-Stage Learning for Surface Roughness Prediction in Ultra-Precision Machining}

\author[P1]{Penghui Ruan}
\ead{penghui.ruan@connect.polyu.hk}
\author[P1]{Divya Saxena}
\ead{divsaxen@comp.polyu.edu.hk}
\author[P1]{Jiannong Cao}
\ead{csjcao@comp.polyu.edu.hk}
\author[P1]{Xiaoyun Liu\corref{cor1}}
\ead{xiaoyun.liu@connect.polyu.hk}
\cortext[cor1]{Corrsponding author.}

\author[P2]{Ruoxin Wang}
\ead{ruoxin.wang@connect.polyu.hk}
\author[P2]{Chi Fai Cheung}
\ead{Benny.Cheung@polyu.edu.hk}

\affiliation[P1]{organization={Department of Computing, The Hong Kong Polytechnic University},
            country={Hong Kong, China}}

\affiliation[P2]{organization={State Key Laboratory of Ultra-precision Machining Technology, Department of Industrial and Systems Engineering, The Hong Kong Polytechnic University},
            country={Hong Kong, China}}

\begin{abstract}
Accurate surface roughness prediction is critical for ensuring high product quality, especially in sectors such as manufacturing, aerospace, and medical devices, where the smallest imperfections can compromise performance or safety. However, this is very challenging due to complex, non-linear interactions among variables, which is further exacerbated when working with limited and imbalanced datasets. Existing methods leveraging traditional machine learning algorithms require extensive domain knowledge for feature engineering and substantial human intervention for model selection. To address these issues, we propose a Neural Architecture Search (NAS)-Driven Multi-Stage Learning Framework, named NASPrecision. This innovative approach autonomously identifies the most suitable features and models for various surface roughness prediction tasks and significantly enhances the performance by multi-stage learning. Our framework operates in three stages: 1) \textbf{architecture search stage}, employing NAS to automatically identify the most effective model architecture; 2) \textbf{initial training stage}, where we train the neural network for initial predictions; 3) \textbf{refinement stage}, where a subsequent model is appended to refine and capture subtle variations overlooked by the initial training stage. In light of limited and imbalanced datasets, we adopt a generative data augmentation technique to balance and generate new data by learning the underlying data distribution. We perform extensive experiments on three distinct real-world datasets, each associated with a different machining technique, and compare with various machine learning algorithms. The experimental results underscore the superiority of our framework, which achieves an average improvement of 18\%, 31\%, and 22\% in terms of Mean Absolute Percentage Error (MAPE), Root Mean Square Error (RMSE), and Standard Deviation (STD), respectively. This significant performance enhancement not only confirms the robustness of our framework but also establishes it as a generic solution for accurate surface roughness prediction. The success of this approach can lead to improved production efficiency and product quality in critical industries while also reducing the need for extensive domain knowledge and human intervention.
\end{abstract}




\end{frontmatter}


\section{Introduction}
Ultra-precision machining (UPM) stands at the forefront of modern manufacturing technologies, instrumental in producing components with extremely high quality at a nanometric surface roughness and a sub-micrometric form accuracy \cite{zhang2015review,li2019machine}. It is pivotal in various industries, including but not limited to optics \cite{li2012design}, electronic and aerospace industries \cite{brinksmeier2012review,yu2012optimized}, where precision and accuracy are paramount. In this highly specialized field, one of the most critical and complex tasks is the prediction of surface roughness. This task is crucial for enhancing production efficiency, minimizing costly trial-and-error iterations intrinsic to machining operations, and ensuring the highest quality of finished products. However, analyzing the machining processes and predicting the surface roughness are non-trivial. The challenge stems from the multifaceted nature \cite{benardos2003predicting} of machining processes, where individual parameters can greatly impact the final results. Furthermore, the intricate and often unclear interactions among these parameters make understanding their collective effect even more challenging.

Traditional machine learning has been identified as a powerful instrument in addressing the complexities involved in predicting surface roughness in the field of UPM. A considerable body of research \cite{salgado2009process,zhang2016effective,kong2020bayesian,ccaydacs2012support,wu2019predictive,li2019prediction,wang2022ensemble} has delved into employing traditional machine learning methodologies to decipher the intricate relationships that exist between the parameters of machining experiments and the resultant surface roughness. Nevertheless, these methods often necessitate considerable human intervention and expertise. Primary among these requirements is the complexity of feature engineering, which requires deep domain knowledge to identify important features relevant to subsequent tasks. Similarly, choosing the right model to accurately represent the relationship between the selected features and the target variable is a challenging task. These challenges highlight the ongoing need for approaches that reduce dependence on extensive human expertise and intervention.

In this regard, we introduce an automatic, universally applicable Neural Architecture Search (NAS)-driven multi-stage learning framework, named NASPrecision for different surface roughness prediction tasks. NASPrecision framework includes three stages, namely, the architecture search stage, the initial training stage, and the refinement stage. The first stage of NASPrecision framework employs NAS \cite{baker2016designing,zoph2016neural}, which automatically identifies the most effective neural network architecture. We further incorporate Bayesian Optimization \cite{snoek2012practical} to speed up the search process given the vast search space associated with NAS. This method substantially minimizes the necessity for manually exploring the vast search space, resulting in a more efficient and precise design and selection process for our machine learning model. Following the architecture search, NASPrecision framework progresses to the initial training stage, where the identified neural network is trained on the dataset to establish an initial predictive model. This stage forms the foundation of NASPrecision framework, setting a baseline for model performance. To rectify potential biases inherent in the architecture search stage and to enhance accuracy for precision-sensitive domains, our model transits to the refinement stage. Here, the initial model is frozen to preserve its fundamental predictive ability, while a secondary, trainable model is appended. This additional model specifically targets high-frequency components in the data distribution and focuses on rectifying residual errors or subtle nuances that may have been missed in the earlier stage. It acts as a sophisticated adaptation mechanism, adding an extra layer of refinement to the prediction process. This approach not only maintains the strengths of the initial model but also significantly improves overall accuracy, ensuring a more nuanced and precise solution. 

Recognizing the constraints of limited and imbalanced datasets in UPM, a field prone to overfitting, NASPrecision framework integrates a generative data augmentation strategy at an early stage. We employ a generative model to closely mimic the original data distribution, enabling us to expand and balance the existing sparse dataset. This improvement in data augmentation not only enhances the robustness of our models but also enables better generalization on unseen data. 

Through comprehensive experiments conducted on three highly diverse datasets, our model demonstrates the impressive capability and robustness of our proposed model. Our proposed NASPrecision framework achieves an average improvement of 18\%, 31\%, and 22\% in terms of MAPE, RMSE, and STD respectively. Despite the marked differences among these datasets, our model consistently outperforms many existing machine learning algorithms, showcasing its wide applicability. Moreover, we delve deeper into the understanding of NASPrecision framework through extensive hyperparameter analysis and ablation studies, which further highlight the efficacy and solid grounding of our approach. 

The paper is structured as follows. Section 2 reviews the related works. Section 3 describes the proposed framework. Section 4 describes the experimental setup. Section 5 performs result analysis and discusses the proposed algorithm.  Section 6 concludes and suggests future research directions.

\section{Related Works}
\subsection{Surface Roughness Prediction Based on Machine Learning}
Over the years, machine learning has garnered significant interest for predicting surface roughness, owing to its robust capabilities to approximate both linear and non-linear functions, thereby effectively addressing the regression problem at hand.

In simpler scenarios, linear models often suffice to encapsulate the underlying relationships. Salgado et al. \cite{salgado2009process}, for example, put forth an in-process estimation method for predicting surface roughness in the turning process utilizing Least-Squares Support Vector Machines (LS-SVM). This model integrated cutting parameters, tool geometry parameters, and vibration signals as inputs. Similarly, Zhang et al. \cite{zhang2016effective} applied LS-SVM to different materials, including AISI4340 steel and AISID2 steel. Another instance is the work by Kong, et al.  \cite{kong2020bayesian}, in which four different Bayesian linear regression models are used to enhance the prediction accuracy of the milling operation.

However, more complex problems often necessitate the use of non-linear models. For instance, \cite{ccaydacs2012support} developed and applied three SVM variants (LS-SVM, Spider SVM, and SVM-KM) to predict the surface roughness of AISI 304 during turning. Wu et al. \cite{wu2019predictive} utilized an array of machine learning techniques, including Random Forests (RFs), Support Vector Regression (SVR), Ridge Regression (RR), and LASSO, to model the surface roughness of additively manufactured parts.

Further strides were made in improving the robustness of such models by employing ensemble methods. Li et al. \cite{li2019prediction} built a data-driven predictive model using a weighted combination of six algorithms (RF, AdaBoost, Classification and Regression Trees (CART), SVR, RR, and Random Vector Functional Link (RVFL) network), with weights computed by the Sequential Quadratic Optimization (SQP) method. In the same vein, ELGA \cite{wang2022ensemble} deployed a genetic algorithm to amalgamate various basic regression algorithms.

Although significant advancements have been achieved in the prediction of surface roughness, the process still requires intricate human intervention and relies on trial-and-error to identify the optimal model. These approaches are not readily applicable when addressing various machining techniques. However, our proposed methods are designed to automate this procedure, providing a generic solution that is adaptable across different machining contexts.

\subsection{Neural Architecture Search}
Neural Architecture Search (NAS) has become an increasingly popular topic in the field of machine learning. Its fundamental aim is to automate the process of designing neural network architectures, thus mitigating the need for extensive expertise and significant time investments typically associated with manual network design \cite{zoph2016neural,baker2016designing}. NAS operates by searching through a predefined space of potential architectures, aiming to find the one that optimizes a given objective function. This function usually pertains to the model's performance on a validation dataset. Over the past few years, NAS methods have outperformed manually designed architecture in many machine learning tasks such as image classification \cite{zoph2018learning, real2019aging}, semantic segmentation \cite{chen2018searching} and language processing \cite{liu2018darts}. Despite the great success of NAS in machine learning tasks, its application in the context of engineering is a relatively uncharted area. 

\begin{figure*}[t!]
    \centering
    \includegraphics[width=1\textwidth]{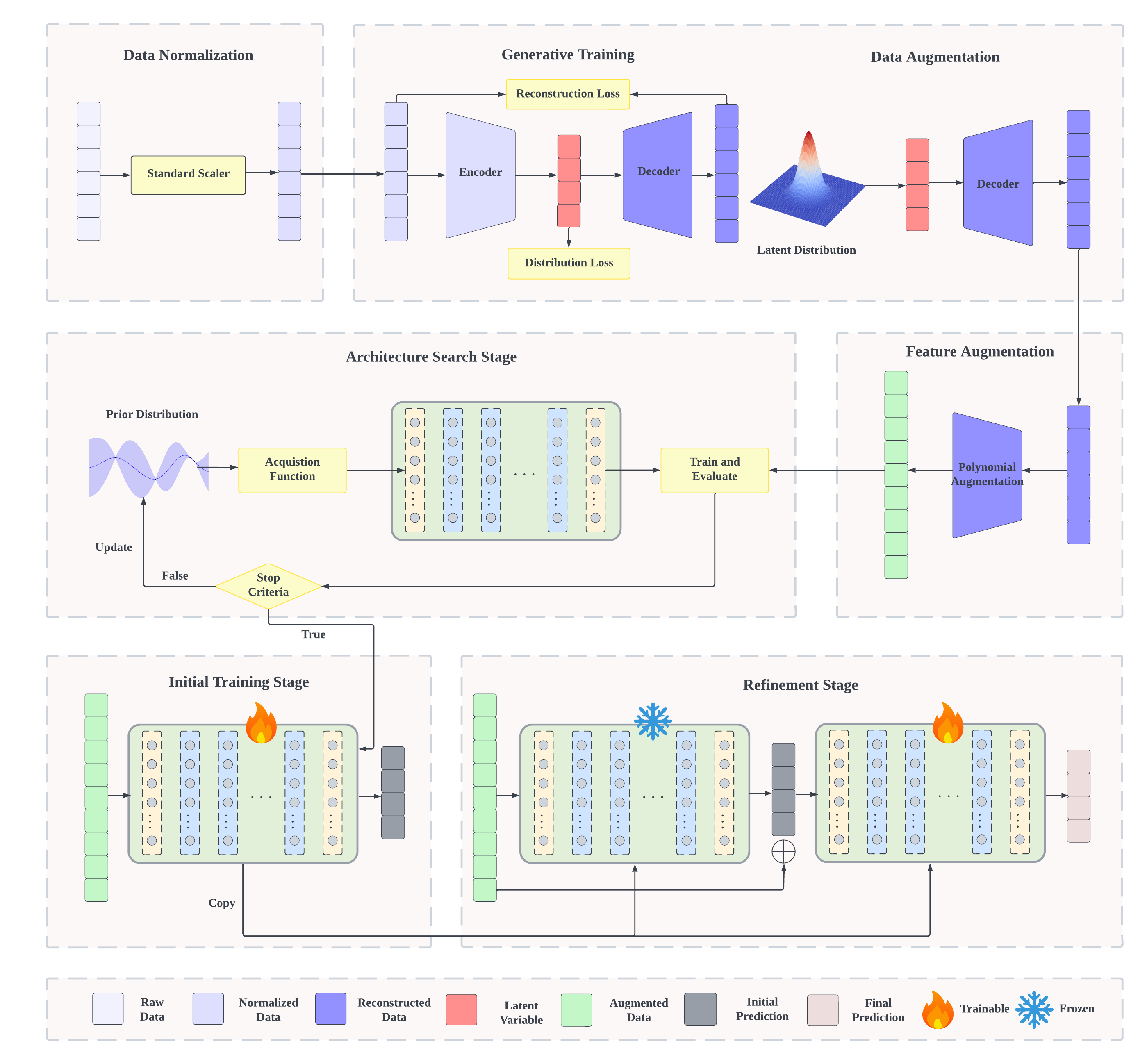}
    \caption{The proposed NAS-driven multi-stage learning framework, NASPrecision}
    \label{fig:architecture}
\end{figure*}
\section{Neural Architecture Search-Driven Multi-Stage Learning}
Figure \ref{fig:architecture} illustrates our proposed multi-stage framework, NASPrecision. The process begins with data normalization, setting the stage for subsequent operations. Following this, a generative model is trained on the normalized data to effectively learn the underlying data distribution. Leveraging the generative model, we then generate new samples from the learned distribution, thereby augmenting and balancing our dataset. This augmented dataset is further processed through feature augmentation, enhancing the feature richness. The next phase involves our architecture search stage, where the enhanced dataset is used to identify the optimal neural network architecture. Once this optimal architecture is determined, we proceed to the initial training stage. Here, the model is trained on the augmented dataset to provide an initial set of predictions. The process culminates in the refinement stage, wherein a secondary trainable model is integrated with the initially trained model (now frozen) to refine and improve the predictions. Algorithm \ref{alg:NAS-Multi-Stage} shows the pseudocode of our proposed algorithm.

\subsection{Data Normalization}
An integral part of our data preprocessing involves data normalization, a crucial step in preparing our dataset for effective model training. Normalization standardizes the range of our data features, ensuring that each feature contributes proportionately to the final prediction and improves the convergence speed during the training process. In this study, we have employed standard normalization, also known as z-score normalization, which adjusts the features so they have the properties of a standard normal distribution with a mean of 0 and a standard deviation of 1. This is mathematically represented as:

\begin{equation}
x_{\text{normalized}} = \frac{x - \mu}{\sigma}
\end{equation}

where $x$ is the original feature value, $\mu$ is the mean of the feature, and $\sigma$ is the standard deviation of the feature. This transformation helps in dealing with the challenges posed by different scales among various features and enhances the model's performance, especially in models sensitive to feature scaling.
\subsection{Generative Data Augmentation}
Generative data augmentation stands out as a potent technique in machine learning for enhancing the efficacy of learning algorithms. It employs advanced generative models, such as Variational Autoencoders (VAEs) \cite{kingma2013auto} and Generative Adversarial Networks (GANs) \cite{goodfellow2020generative}, or Diffusion Models \cite{sohl2015deep,song2020score,ho2020denoising} to grasp and replicate the underlying distribution of existing datasets. This approach is instrumental in generating new, synthetic data samples, thereby artificially expanding the dataset. One of the critical advantages of this technique is its ability to mitigate issues of data imbalance. In this step, we employ a VAE for generative data augmentation. The VAE primarily focuses on data reconstruction, achieved through a sophisticated encoding-decoding process that involves learning and sampling from latent space representations. It begins by encoding input data $x$ into a latent variable $z$ using an encoding network $q_{\phi}(z|x)$. This encoding process learns a distribution over the latent variables, where a Gaussian distribution is typically assumed. From this latent space, the VAE can then generate new data by sampling points from the learned latent distribution and feeding these points into a decoding network $p_{\theta}(z|x)$. This network reconstructs data that is similar to the original input but with variations introduced by the sampling process. The reconstructed data thus augment the original dataset, providing additional variability that can be beneficial for training robust machine learning models. By leveraging the VAE's ability to create diverse and representative data instances, we enhance the dataset's richness and improve model generalization.
The VAE is trained by maximizing the Evidence Lower Bound (ELBO) given as follows:
\begin{equation}
    ELBO = \mathbb{E}_{q_{\phi}(z|x)}[log p_{\theta}(x|z)] - \mathbb{D}_{KL}(q_{\phi}(z|x)||p(z))
\end{equation}
\begin{equation}
    \mathbb{D}_{KL}(q_{\phi}(z|x)||p(z)) = \mathbb{E}_{q_{\phi}(z|x)}log\frac{q_{\phi}(z|x)}{p(z)}
\end{equation}

where $p(z), p_{\theta}(x|z), q_{\phi}(z|x)$ are prior, likelihood, and posterior, respectively. The first part of ELBO is essentially the reconstruction loss, quantifying the discrepancy between the reconstructed and original samples. The second part, the distribution (or regularization) loss, calculates the KL-divergence of the learned latent space distribution $q_{\phi}(z|x)||p(z)$ from a Gaussian prior $p(z)$, thereby encouraging a Gaussian structure in the latent space. After the training, we can generate data with the trained VAE, which can be written as $z \sim p(z)$, and $x_{new} \sim p_{\theta}(x|z)$.
    

\subsection{Feature Augmentation}
The underlying relationships between the parameters of ultra-precision machining and final surface roughness are complicated. The raw features may not be sufficient to capture all the relevant information. To address this, we enhance the feature set by applying a polynomial transformation to the raw features. This process can be mathematically articulated as:
\begin{equation}
    \phi^p(\mathbf{x}) = \left[ 1, x_1, \ldots, x_n, x_1^2, x_1x_2, \ldots, x_{n-1}x_n, x_n^2, \ldots, x_n^p \right]^T
\end{equation}
where $\mathbf{x}=(x_1,x_2,...,x_n)$, $n$ is the dimensionality of the input and $p$ is the order of polynomial augmentation.

\subsection{Architecture Search Stage}
\begin{figure*}[t]
\centering
\includegraphics[width=\textwidth]{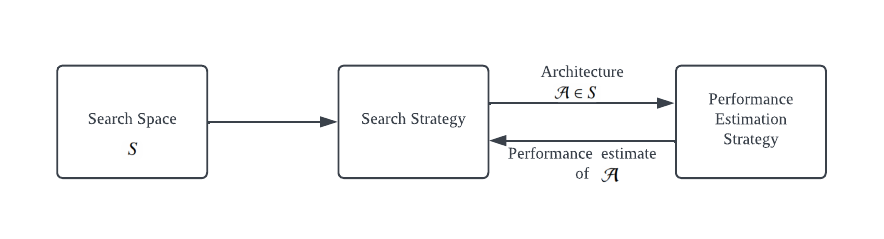}
\caption{Abstract illustration of Neural Architecture Search methods \cite{elsken2019neural}. A search strategy selects an architecture $\mathcal{A}$ from a predefined search space $S$. The architecture is passed to a performance estimation strategy, which returns the estimated performance of A to the search strategy.}
\label{fig:NAS}
\end{figure*}

In the architecture search stage, we perform a NAS to find the most suitable architecture for the given problem. The NAS normally consists of three components: search space, search strategy, and performance estimate strategy, as shown in Figure \ref{fig:NAS}. The search space $S$ delineates the entirety of potential model architectures. The search strategy outlines the methodology employed to traverse this space, aiming to identify the most promising architectural candidate $\mathcal{A}$. Subsequently, the performance estimation strategy is tasked with providing a performance evaluation for the selected candidate $\mathcal{A}$.
In the subsequent sections, we will detail the specific design choices made in our architecture search stage.

\subsubsection{Search Space}
The design of the search space plays a pivotal role in influencing the performance of the model. As per the Universal Approximation Theorem \cite{hornik1989multilayer,csaji2001approximation}, a neural network possesses the capability to approximate any function, given an appropriate architecture. In light of this, we design of search space as a composite of fundamental neural network building blocks. This includes the number of hidden layers, the number of neurons per hidden layer, the choice of activation functions, batch size, learning rate, and loss function.

\begin{figure*}[htbp]
\centering
 \includegraphics[width=\textwidth]{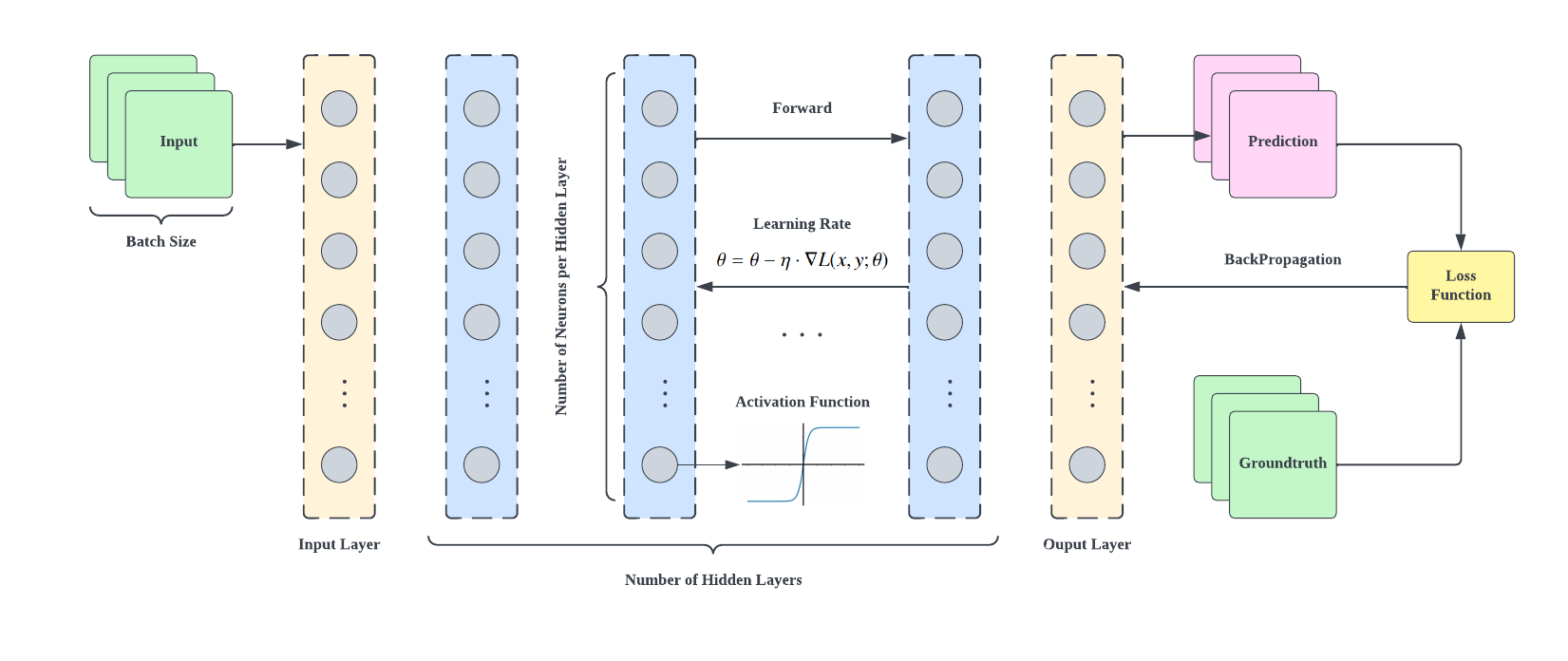}
\caption{Semantic meanings of the Search Space}
\label{fig:NN}
\end{figure*}

To fully capture the complex and diverse relationship, we design our search space to include both linear and non-linear activation functions. This approach is intended to provide the necessary flexibility and adaptability in the model design, catering to the varying complexities of the data patterns. The detailed structure of our search space is enumerated in Table \ref{table:search space}, and the semantic implications of each element within the search space are illustrated in Figure \ref{fig:NN}.

\subsubsection{Search Strategy}
The most straightforward search strategy is the brute-force approach. However, due to the relatively large search space of NAS, brute-force methods often prove infeasible. To more effectively explore and exploit the search space, we leverage Bayesian optimization \cite{snoek2012practical}, a powerful strategy for global optimization of black-box functions. Bayesian optimization employs a probabilistic model, typically a Gaussian Process (GP) \cite{williams1995gaussian}, to represent a prior distribution over the unknown objective function, encapsulating our beliefs about its behavior based on prior evaluations.

Given a search space $S$, each architecture $\mathcal{A} \in S$ is represented as a hyperparameter vector. The performance of an architecture is given by a function $g(\mathcal{A})$. We model our belief about $g$ using a GP with mean function $m(\mathcal{A})$ and covariance function $k(\mathcal{A}, \mathcal{A}')$.

After evaluating $g(\mathcal{A})$ for a set of architectures, we update our GP. The mean and variance of the GP at any architecture $\mathcal{A}$ are updated using the formula:
    \begin{align}
m(\mathcal{A}) &= k(\mathcal{A}, \mathbf{A})K^{-1}\mathbf{Y} \\
v(\mathcal{A}) &= k(\mathcal{A}, \mathcal{A}) - k(\mathcal{A}, \mathbf{A})K^{-1}k(\mathbf{A}, \mathcal{A})
\end{align}

where $\mathbf{A}$ is the matrix of evaluated architectures, $\mathbf{Y}$ is the vector of corresponding performances, $K$ is the covariance matrix with elements $K_{ij} = k(\mathbf{A}_i, \mathbf{A}_j)$, and $k(\mathcal{A}, \mathbf{A})$ is the vector of covariances between $\mathcal{A}$ and each architecture in $\mathbf{A}$.

The acquisition function $\alpha(\mathcal{A})$, which directs the search strategy is then computed for each $\mathcal{A} \in S$. Different variants of the acquisition function balance the exploitation and exploration differently. For example, the Expected Improvement (EI) criterion balances exploitation and exploration by preferring regions where the model predicts the potential for significant improvement. 
\begin{equation}
\alpha_{EI}(\mathcal{A}) = max(g(\mathcal{A}) - g(\mathcal{A^*}),0)
\end{equation}

On the other hand, the Upper Confidence Bound (UCB) criterion primarily encourages exploration by favoring architectures where the predictive uncertainty is high.
\begin{equation}
\alpha_{UCB}(\mathcal{A}) = m(\mathcal{A}) + \beta \sigma(\mathcal{A})
\end{equation}
where $\beta$ is a tradeoff parameter, and $\sigma(\mathcal{A})=\sqrt{K(\mathcal{A},\mathcal{A})}$ is the marginal standard deviation.

We then select the architecture that maximizes the acquisition function, evaluate $g$ at that architecture, update the GP, and repeat until a stopping condition is met.
\subsubsection{Performance Estimation Strategy}
In our design, the strategy for evaluating performance is simple and effective. For each neural network architecture $\mathcal{A}$ that we consider, we first train it on our training dataset. This step allows the network to learn and adapt to the specific patterns and features of our data. After training, we evaluate the model's performance on a separate validation set. This validation set is different from the training data, ensuring that we are testing how well the network can generalize to new, unseen data. We measure its performance using the loss function. This strategy aims to find the architecture $\mathcal{A^*}$ that not only learns well from the training data but also performs well on the validation data, showing good generalization on unseen data.

\subsection{Initial Training Stage}
Upon finalizing the architecture search, we proceed to the initial training stage, where the model is trained on the dataset. This stage is crucial for establishing a baseline performance of the model. This process can be mathematically represented as follows:
\begin{equation}
\theta_1^* = \arg\min_{\theta_1} \mathbb{E}_{(x, y) \sim \mathcal{D}} [L(f_{\mathcal{A^*}}(x;\theta_1), y)]
\end{equation}
In this equation, $\theta_1$ denotes the set of parameters for the neural network characterized by the optimal architecture $\mathcal{A^*}$, identified in the previous architecture search phase. The goal of this stage is to find the parameter set $\theta_1^*$ that minimizes this expected loss, thereby calibrating the model to capture the primary relationships and patterns present in the data.

\begin{algorithm}[!b]
\caption{Neural Architecture Search-Driven Multi-Stage Learning}
\begin{algorithmic}[1]
\Require{Dataset $D$, Search space $S$, Learning rate $\eta$, Acquisition function $\alpha$}
\Ensure{Best Model $f_{\mathcal{A^*}}(x;\theta_1^*,\theta_2^*)$}
\State Initialize Gaussian Process
\While{stopping criterion for NAS not met}
    \State Select architecture $\mathcal{A}$ maximizing acquisition function $\alpha(\mathcal{A})$
    \State Train and evaluate $g(\mathcal{A})$ 
    \State Update Gaussian Process with $\mathcal{A}, g(\mathcal{A})$
\EndWhile
\State Initialize $f_{\mathcal{A^*}}(;\theta_1)$ with best architecture $\mathcal{A}^*$, and random parameters $\theta_1$
\While{stopping criterion for $f_{\mathcal{A^*}}(;\theta_1)$ not met}
    \State Compute $L(f_{\mathcal{A^*}}(x;\theta_1),y)$
    \State Update $\theta_1= \theta_1 - \eta \nabla_{\theta_1}L(f_{\mathcal{A^*}}(x;\theta_1),y)$
\EndWhile

\State Initialize $f_{\mathcal{A^*}}(x;\theta_2)$ with same architecture $\mathcal{A}^*$, and the parameters $\theta_2 = \theta_1^*$
\State Freeze $f_{\mathcal{A^*}}(x;\theta_1)$
\While{stopping criterion for $f_{\mathcal{A^*}}(x;\theta_2)$  not met}
    \State For each training example $x$, compute $f_{\mathcal{A^*}}(x;\theta_1^*)$
    \State Compute $L(f_{\mathcal{A^*}}(x,f_{\mathcal{A^*}}(x;\theta_1^*);\theta_2),y)$
    \State Update $\theta_2= \theta_2 - \eta \nabla_{\theta_2}L(f_{\mathcal{A^*}}(x,f_{\mathcal{A^*}}(x;\theta_1^*);\theta_2),y)$
\EndWhile
\State Output model $f_{\mathcal{A^*}}(x;\theta_1^*,\theta_2^*)=f_{\mathcal{A^*}}(x,f_{\mathcal{A^*}}(x;\theta_1^*);\theta_2^*)$
\end{algorithmic}
\label{alg:NAS-Multi-Stage}
\end{algorithm}

\subsection{Refinement Stage}
The initial model serves as a foundational model in NASPrecision framework, providing an initial approximation of the surface roughness. While effective in capturing the broader patterns, this model inherently incorporates certain biases resulting from the NAS process and may overlook finer details and variations in the data. To address these limitations, we strategically freeze the initial model, preserving its broad approximations. Subsequently, we append an additional trainable model with the same architecture to the frozen initial model. This additional model takes the original input and the primary prediction given by the initial model. It is specifically designed to capture and rectify the finer nuances and biases that the initial model might have missed. Functioning as a refinement stage, this step significantly enhances the model's capacity to discern subtleties, thereby boosting overall performance.
The refinement stage is mathematically articulated as follows:
\begin{equation}
\theta_2^* = \arg\min_{\theta_2} \mathbb{E}_{(x, y) \sim D}[L(f_{\mathcal{A^*}}(x,f_{\mathcal{A^*}}(x;\theta_1^*);\theta_2), y)]
\end{equation}

where $\theta_2$ is the parameters of the additional model.

\section{Experiments}

\subsection{Evaluation Metric}
In assessing the performance of our proposed model against baseline models, we employ three key metrics: Root Mean Squared Error (RMSE), Mean Absolute Percentage Error (MAPE), and the Standard Deviation (STD) of the prediction error. RMSE and MAPE are pivotal in quantifying the discrepancies between actual and predicted values. RMSE is particularly adept at highlighting the impact of outliers, as it disproportionately emphasizes larger errors. On the other hand, MAPE provides a percentage-based perspective of the average error, making it straightforward to interpret. Additionally, we use STD as a measure to evaluate the spread of prediction errors. A lower STD indicates that the predicted values are more closely clustered around the mean, suggesting a higher level of consistency in the predictions. This approach ensures a comprehensive assessment, taking into account not only the average accuracy but also the variability and outlier sensitivity of the predictions.

\begin{equation}
    MAPE = \frac{1}{N}\underset{n}{\sum}|\frac{y_n-\hat{y}_n}{y_n}|
\end{equation}

\begin{equation}
    RMSE = \sqrt{\frac{1}{N}\underset{n}{\sum}(y_n-\hat{y}_n)^2}
\end{equation}
\begin{equation}
    STD = \sqrt{\frac{1}{N}\underset{n}{\sum}(y_n - \hat{y_n}-\Bar{e})^2} \; s.t. \; \Bar{e} = \frac{1}{N}\underset{n}{\sum}{y_n - \hat{y_n}}
\end{equation}
\subsection{Dataset}
In the experiments, we use three surface roughness prediction datasets to evaluate the proposed algorithms. 

\subsubsection{MJP Dataset}
In the MJP experiments \cite{wang2022ensemble}, the 3D-printed 316L stainless steel components were polished with different process parameters by a ZEEKO IRP200 machine. The detailed parameters are shown in Table \ref{table:MJP}. 
\begin{table}[htbp]
\centering
\caption{Parameter setting of MJP}
\label{table:MJP}
{
\begin{tabular}{ll}
\toprule
Polishing parameter & Range \\ \midrule
Feed rate($f$) & 10,15,20,25,30,40,60,80 mm/min \\ 
Fluid pressure($P$) & 4,5,6,7,8,9,10 bar \\ 
Tool offset($TO$) & 2.5,5,7.5,10,12.5,15 mm \\ 
Step distance($d$) & 0.1,0.2,0.3,0.4,0.5,0.6,0.7,0.8 mm \\ 
Surface direction & TS (top surface) \\
\bottomrule
\end{tabular}
}
\end{table}

\subsubsection{CNC Turning Dataset}
The CNC turning dataset \cite{ccaydacs2012support} was collected from a series of turning experiments. These experiments utilized a JOHNFORD TC-35 lathe machine equipped with a Fanuc18-T CNC control, a programmable tailstock, and a 15 kW drive motor, enabling a maximum spindle speed of 3,500 rpm. Cutting tools employed in these trials were commercial-grade cemented carbide inserts provided by Kennametal with the geometry of CNMG 120408. The workpieces used for the experiments were made of AISI 304 austenitic stainless steel which is widely used in aircraft fittings, and aerospace components for severe chemical environments \cite{xavior2009determining}. The dataset contains 27 different combinations of turning parameters with a three-level full factorial experimental design. The parameters and their factor levels of the experiments are shown in Table \ref{table:CNC}.
\begin{table}[htbp]
\centering
\caption{Parameter setting of CNC turning}
\label{table:CNC}
{
\begin{tabular}{p{.2\textwidth}p{.06\textwidth}p{.06\textwidth}p{.06\textwidth}}
\toprule
Turning parameter & Level 1 & Level 2 & Level 3 \\ \midrule
Cutting speed (m/min)& 30 & 60 & 90 \\ 
Feed rate (mm/rev) & 0.15 & 0.25 & 0.35 \\
Depth of cut (mm) & 0.5 & 1 & 1.5 \\
\bottomrule
\end{tabular}
}
\end{table}

\subsubsection{Cutting Vibration Dataset}
The cutting vibration dataset \cite{salgado2009process} under consideration has been gathered from a series of turning experiments conducted using AISI 8620 steel as the workpiece material. The experimentation involved various ISO types of TiN-coated carbide inserts, specifically CCMT 120404, CCMT 120408, TCMT 110204, VCMT 160404, and VCMT 160408. They used the accelerometers (Kistler type 8742A50 and Kistler 5807 A amplifiers) to measure the cutting vibrations. This dataset provides the vibration features extracted from cutting vibrations recorded during the turning operations, providing a rich and detailed source of information for the analysis of such processes. Table \ref{table:Cut} lists the parameters of this dataset.
\begin{table}[htbp]
\centering
\caption{Parameters of cutting vibration dataset}
\label{table:Cut}
{
\begin{tabular}{p{.2\textwidth}p{.2\textwidth}}
\toprule
Cutting Parameter & Meaning \\ \midrule
$v_c$ (m/min) & Cutting speed \\ 
$f$ (mm/rev) & Feed rate  \\ 
$d$ (mm) & Depth of cut  \\ 
$r$ (mm) & Nose radius  \\ 
$A$ & Nose angle  \\ 
$V_B$ ($\mu$m) & Tool flank wear  \\
\bottomrule
\end{tabular}
}
\end{table}

\begin{table*}[htbp]
\centering
\caption{Search Space}
\label{table:search space}
{
\begin{tabular}{p{.46\textwidth}p{.46\textwidth}}
\toprule
Architecture & Range \\ \midrule
Number of Hidden Layers & 1, 2, ..., 10  \\ 
Number of Neurons per Hidden Layer&  10, 11, ..., 100\\
Activation Functions & ReLU, Tanh, Identity, ELU, LeakyReLU, Sigmoid \\ 
Batch Size & 4, 8, 16, 32, 64\\ 
Learning Rate &   [0.0001,0.05]     \\ 
Loss Function & L1, L2 \\
\bottomrule
\end{tabular}
}
\end{table*}

\subsection{Baselines}
In order to substantiate the superiority of our proposed methods, we undertake a comprehensive comparison spanning three distinct classes of algorithms. These classes encompass linear methodologies, non-linear methodologies, and ensemble methodologies.
\subsubsection{Linear Methods}
\textbf{Linear Regression} is a linear approach to model the linear relationship between a dependent variable and one or more independent variables. For a dataset with input matrix $X$, the prediction $\hat{Y}$ is:

\begin{equation}
\hat{Y} = Xw
\end{equation}
Where $w$ is the weight matrix for linear regression. The linear regression then finds the weight matrix by minimizing the following objective function:
\begin{equation}
    f(w) = \frac{1}{N}||Y-Xw||^2_2 
\end{equation}

\textbf{Least Absolute Shrinkage and Selection Operator (LASSO) \cite{tibshirani1996regression} } is a linear regression technique that incorporates L1 regularization. Regularization adds a penalty term to the objective function, which helps to shrink the coefficients of less important features towards zero. Mathematically, LASSO aims to minimize the following objective function:
\begin{equation}
    f(w) = \frac{1}{N}||Y-Xw||^2_2 + \lambda_1||w||_1
\end{equation}
where $\lambda_1$ is the learnable Lagrangian parameter.

\textbf{Ridge Regression (RR) \cite{hoerl1970ridge}} is another linear regression technique that uses L2 regularization. Similar to LASSO, RR adds a penalty term to the objective function to control model complexity and mitigate the effects of multicollinearity.
Mathematically, RR aims to minimize the following objective function:
\begin{equation}
    f(w) = \frac{1}{N}||Y-Xw||^2_2 + \lambda_2||w||^2_2
\end{equation}
where $\lambda_2$ is the learnable Lagrangian parameter of RR.

\textbf{Elastic Net Regression (ENR) \cite{zou2005regularization}} is a powerful regularization technique that combines the strengths of both LASSO and RR to improve the performance of linear regression models. The Elastic Net technique balances the L1 and L2 penalties in the objective function, which can be represented mathematically as:
\begin{equation}
    f(w) = \frac{1}{N}||Y-Xw||^2_2 + \alpha \lambda||w||^2_1 +  (1-\alpha)\lambda||w||^2_2 
\end{equation}
where $\lambda$ and $\alpha$ are learnable parameters.

\textbf{Linear Support Vector Regression (SVR-Linear) \cite{drucker1996support}} is an extension of Support Vector Machines (SVM) \cite{cortes1995support} to regression problems. The aim of SVR is to find a function $f(x)$ that has at most $\varepsilon$ deviation from the actually obtained targets $y_i$ for all the training data and, at the same time, is as flat as possible. Formally, given a set of training examples $(x_{i}, y_{i})$ where $x_{i} \in R^{n}$ is a feature vector and $y_{i} \in R$ is the target, SVR solves the following optimization problem:
\begin{align}
& \underset{w,b,\xi,\xi^*}{min} & \frac{1}{2}||w||^{2} + C \sum_{i=1}^{N}(\xi_i + \xi_i^*) \\
& s.t. & y_{i} - w \cdot x_{i} - b \leq \varepsilon + \xi_i, \\
& & w \cdot x_{i} + b - y_{i} \leq \varepsilon + \xi_i^*, \\
& & \xi_i, \xi_i^* \geq 0, ; i = 1, \ldots, N.
\end{align}

Here, $w$ is the weight vector, $b$ is the bias, $\xi_i$ and $\xi_i^*$ are slack variables that allow for points outside the $\varepsilon$-insensitive tube, and $C$ is the regularization parameter.

\subsubsection{Non-Linear Methods}
\textbf{K-Nearest Neighbor Regression (KNNR) \cite{fix1985discriminatory}} is a non-parametric method that predicts the output of a new instance based on the outputs of its k-nearest neighbors in the training set. For a given instance $x$, the prediction $\hat{y}$ is:
\begin{equation}
\hat{y} = \frac{1}{k}\sum_{i\in N_k(x)}y_i
\end{equation}
where $N_k(x)$ is the set of the k-nearest neighbors of $x$ and $y_i$ is the output of the $i^{th}$ neighbor.

\textbf{Gaussian Process Regression (GPR)} is a Bayesian, non-parametric method used for regression. In GPR, the prediction for a new input is made by taking a weighted average of the outputs of the observed data. Additionally, each prediction is associated with a Gaussian distribution, which is described by a mean function and a covariance function (or kernel). The kernel function represents the similarity between different data points. Mathematically, for a new test point $x_{*}$, the predictive distribution given training data $(X, y)$ is given by:
\begin{equation}
p(y_{*} | x_{*}, X, y) = \mathcal{N}(y_{*} | K_{*T} K^{-1}y, K_{**} - K_{*T} K^{-1} K_{*})
\end{equation}
Here, $K$ is the kernel matrix of the training data points, $K_{*}$ is the kernel evaluations between the test point and training points, and $K_{**}$ is the kernel evaluation at the test point. 

\textbf{SVR-RBF (Radial Basis Function)} is a version of SVR that uses RBF as kernel function to transform data into high dimensional space where the data is separable. The decision function for SVR-RBF is:
\begin{equation}
f(x) = \sum_{i=1}^{N} (\alpha_i - \alpha_i^*) K(x, x_i) + b
\end{equation}
\begin{equation}
K(x_i, x_j) = \exp\left(-\frac{|x_i-x_j|^2}{2\sigma^2}\right)
\end{equation}
where $\alpha_i$ and $\alpha_i^*$ are the dual coefficients, $K(x, x_i)$ is the RBF kernel, and $b$ is the bias term.
\subsubsection{Ensemble Methods}
\textbf{Ensemble Learning-Average (EL-Avg)} combines multiple individual models together to produce a final output. When using averaging as a strategy, the final prediction is typically the average of the predictions made by each individual model.

For an ensemble model with $N$ models, and given an instance $x$, the prediction $\hat{y}$ is:
\begin{equation}
\hat{y} = \frac{1}{N}\sum_{i=1}^{N}\hat{y_i}
\end{equation}
where $\hat{y_i}$ is the prediction of ith model.

\textbf{Random Forest (RF) \cite{breiman2001random}} is a powerful, ensemble-based machine learning algorithm that leverages the concept of bagging. It constructs a multitude of decision trees during training and outputs the mean prediction of individual trees for the final prediction. The Random Forest algorithm can be mathematically described using the following equation:
\begin{equation}
    \hat{y} = \frac{1}{N}\sum f_k(x)
\end{equation}
where $f_k(x)$ is the prediction for k-th decision tree given $x$.
\begin{table*}[htbp]
\centering
\caption{Performance Comparison. The best results across each metric are highlighted in bold, while the second-best results are indicated with an underline}
\label{table:performance}
\begin{tabular}{cc|ccc|ccc|ccc}
\toprule
\multicolumn{2}{c|}{Algorithm} & \multicolumn{3}{c|}{MJP} &\multicolumn{3}{c|}{CNC Turning} &\multicolumn{3}{c}{Cutting Vibration}\\ 
 & & MAPE &  RMSE & STD & MAPE &  RMSE & STD & MAPE &  RMSE & STD\\ \midrule
\multirow{5}{*}{Linear}
&LR &0.1454 & 22.5351 & 20.7490             & 0.3984 & 1.1371 & 1.0259            & 0.1653 & 0.6527 & 0.5722\\ 
&RR & 0.1434 & 15.3770 & 14.2775            & 0.2782 & 0.7996 & 0.7517           & \underline{0.0083} & 0.0317 & 0.0267\\ 
&LASSO & 0.1408 & 12.9251 & 12.8836         & 0.2995 & 0.8686 & 0.8368            & 0.0113 & 0.0311 & 0.0208\\ 
&ENR & 0.1380 & 12.4517 & 11.9188           & 0.2765 & 0.8076 & 0.7807            & 0.0109 & \underline{0.0270} & \underline{0.0172}\\ 
&SVR-Linear &  0.2185 & 24.5974 & 22.6432    & 0.2995 & 0.8686 & 0.8368            & 0.0335 & 0.0579 & 0.0551\\ 

&&&&&&&&&& \\

\multirow{3}{*}{Non-Linear}
&KNNR & 0.2196 & 41.1345 & 35.6277          & 0.2525 & 0.7540 & 0.7537            & 0.2118 & 0.5264 & 0.5216\\
&GPR & 0.6365 & 118.4436 & 101.4030         & 0.4886 & 1.3657 & 1.2695            & 0.7879 & 2.1699 & 1.3965\\
&SVR-RBF &0.5362 & 74.8418 & 66.6977        & \underline{0.2235} & \underline{0.7126} & \underline{0.6621}            & 0.1997 & 0.8013 & 0.7760 \\ 
&&&&&&&&&&\\
\multirow{4}{*}{Ensemble}
&RF & 0.2715 & 33.5520 & 32.8778            & 0.4319 & 1.2113 & 1.1545            & 0.0848 & 0.5090 & 0.4670\\ 
&XGBoost & 0.3630 & 32.9600 & 32.1636       & 0.4884 & 1.3650 & 1.2690            & 0.0575 & 0.1227 & 0.1205\\
&EL-Avg & 0.1793 & 13.3310 & 13.1529        & 0.3549 & 0.9914 & 0.9396            & 0.0295 & 0.1040 & 0.0922 \\
&ELGA & \underline{0.1366} & \underline{11.8806} & \underline{11.5650}          & 0.3058 & 0.8724 & 0.8373            & 0.0558 & 0.3155 & 0.2875\\
&&&&&&&&&&\\

Multi-Stage&\textbf{NASPrecision (Ours)}
& \textbf{0.1172} & \textbf{11.4777} & \textbf{9.2018}                & \textbf{0.1403} & \textbf{0.3937} & \textbf{0.4680}            & \textbf{0.0081} & \textbf{0.0146} & \textbf{0.0144}\\ 
\bottomrule
\end{tabular}
\end{table*}

\textbf{eXtreme Gradient Boost (XGBoost) \cite{chen2016xgboost}} is an advanced and efficient implementation of the gradient boosting algorithm, which is widely used for classification and regression tasks in machine learning. The idea is to iteratively train new models to correct the errors made by the previous models, ultimately yielding a model with improved accuracy and generalization capabilities. Mathematically, XGBoost aims to minimize the following objective function:
    
\begin{align}
      & \mathcal{L}(\phi) = \underset{i}{\sum}l(\hat{y_i},y_i) + \underset{k}{\sum}\Omega(f_k) \\
& where \quad \Omega(f) = \gamma T + \frac{1}{2}\lambda||w||^2
\end{align}
Here, $l(\hat{y_i},y_i)$ is the loss function, and $\Omega(f_k)$ is the regularization term for the k-th tree. The prediction is given by the following equation:
\begin{equation}
    \hat{y_i} = \phi(x_i) = \underset{k}{\sum}f_k(x_i), \quad f_k \in \mathcal{F}
\end{equation}
where $\mathcal{F}=\{f(x)=w_{q_(x)}\}(q: \mathcal{R}^m \longrightarrow T,w \in \mathcal{R}^T)$ is the space of regression trees.

\textbf{Ensemble Learning with Genetic Algorithm (ELGA) \cite{wang2022ensemble}} is an ensemble learning algorithm with multiple regression algorithms combined by a genetic algorithm. It contains three modules, namely, the multi-algorithm regression module, the GA module, and the ensemble module. Five basic regression models are combined by GA \cite{holland1992genetic}, and the final prediction is made by weighted combination. Mathematically, ELGA can be formulated as follows:
\begin{align}
    &\underset{\alpha}{min} & \sum_{i=1}^N \sum_{k=1}^K\alpha_kf_k(x_i) -y_i\\
    &s.t.  &\sum_{k=1}^K\alpha_k = 1
\end{align}
where $f_k$ is the regression algorithm, and $\alpha_k$ is the corresponding weight found by GA.

\subsection{Experiment Settings}
Our experimental procedure involved partitioning the dataset into training and testing subsets, maintaining a 9:1 ratio. We implemented a VAE with a three-layer structure, employing ReLU as the activation function for both the encoder and decoder. The dimensions for the hidden layer and the latent space were set at 40 and 20, respectively. The VAE was trained using MSE as the reconstruction loss and KL divergence as the distribution loss. The learning rate was set at 0.0001, and we utilized a batch size of 8 with the Adam \cite{kingma2014adam} optimizer for 200 epochs. Data augmentation was performed by generating data 20 times the size of the original dataset.

We also incorporated second-order polynomial feature augmentation in our approach. The Bayesian optimization was performed using GP\_HEDGE as the acquisition function for the MJP and Cutting Vibration datasets. LCB is used for the CNC Turning dataset. This entailed querying and updating the GP model for 30 times.

When applying ELGA, we adhered strictly to the configuration used in previous work \cite{wang2022ensemble}. The setup included a population size of 10,000 and a maximum of 100 generations, alongside roulette selection, two-point crossover (with a probability of 0.9), and random mutation (with a probability of 0.001).

\section{Experiment Results and Analysis}
The organization of our experimental evaluation spans three distinct sections. In Section 5.1, we benchmark our proposed NASPrecision framework against existing machine learning and ensemble learning algorithms, demonstrating its superiority through comparative analysis. In Section 5.2, we delve into a thorough hyperparameter analysis, which primarily explores the influence of different order polynomial feature augmentation and different acquisition functions. Moving forward to Section 5.3, we undertake an ablation study with a particular focus on the third stage of proposed NASPrecision framework. Lastly, in Section 5.4, we lay out some of the constraints and limitations inherent in NASPrecision. 
\begin{table*}[t]
\centering
\caption{Corresponding Architecture of Architecture Search Stage}
\label{table:architecture}
{
\begin{tabular}{p{.4\textwidth}p{.16\textwidth}p{.16\textwidth}p{.16\textwidth}}
\toprule
Architecture & MJP & CNC Turning & Cutting Vibration \\ \midrule
Number of Hidden Layers &  1 & 6 & 7 \\
Number of Neurons per Hidden Layer&  10 & 51 & 14 \\
Activation Functions & LeakyReLU & Sigmoid & ELU \\
Batch Size & 16 & 4 & 64 \\
Learning Rate & 0.0449 & 0.0029 & 0.0021 \\ 
Loss Function & L1 & L1 & L2 \\
\bottomrule
\end{tabular}
}
\end{table*}
\begin{table*}[t]
\centering
\caption{Ablation Study on Refinement Stage}
\label{table:ablation}
\begin{tabular}{cccccccc}
\toprule
\multicolumn{2}{c}{Dataset}&\multicolumn{2}{c}{MAPE} & \multicolumn{2}{c}{RMSE} & \multicolumn{2}{c}{STD}\\ 
 & & w/o Third Stage & w  Third Stage &w/o Third Stage&w Third Stage &w/o Third Stage&w Third Stage  \\ \midrule
\multicolumn{2}{c}{MJP} & 0.1280 & 0.1172 & 16.1367 & 11.4777 & 13.7623 & 9.2018 \\
\multicolumn{2}{c}{CNC Turning} & 0.1474 & 0.1403 & 0.4227 & 0.3937 & 0.5174 & 0.4680   \\
\multicolumn{2}{c}{Cutting Vibration} & 0.0388 & 0.0081 & 0.0946 & 0.0146 & 0.1011 & 0.0144 \\
\bottomrule
\end{tabular}
\end{table*}

\subsection{Cross-Dataset Evaluation and Performance Analysis}
In this section, we undertake a comprehensive evaluation of our proposed methods, conducting experiments across three distinct datasets. Our methods demonstrate their superior performance by outperforming all other considered algorithms with relatively significant margins across all three datasets, as shown in Table \ref{table:performance}. Specifically, we achieve an average improvement of 18\%, 31\%, and 22\% in terms of MAPE, RMSE, and STD, respectively.  The ensuing results from these experiments provide compelling evidence of the effectiveness of our approach. The corresponding architecture for each dataset is listed in Table \ref{table:architecture}.

\subsection{Hyperparameter Analysis}
In this section, we conduct an in-depth analysis of the fundamental hyperparameters of our proposed algorithm, with a focus on the order of polynomial feature augmentation and the selection of acquisition functions in Bayesian optimization.

\begin{figure}[!p]
    \centering
    \includegraphics[width=.35\textwidth]{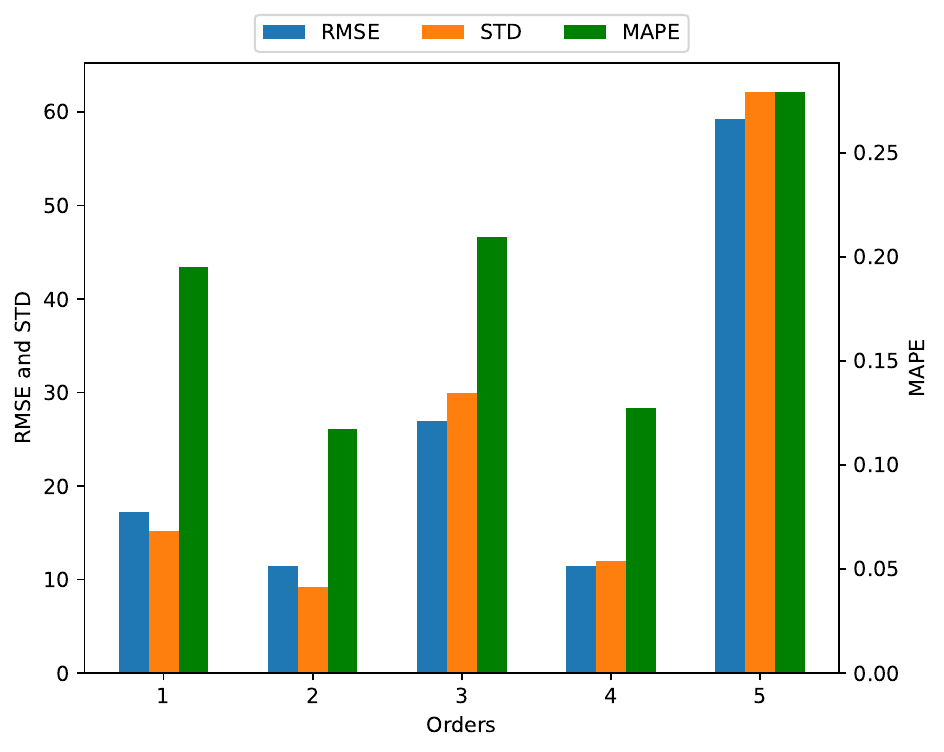}
    \caption{Different order of feature augmentation for MJP dataset}
    \label{fig:order-MJP}
\end{figure}

\begin{figure}[!p]
    \centering
    \includegraphics[width=.35\textwidth]{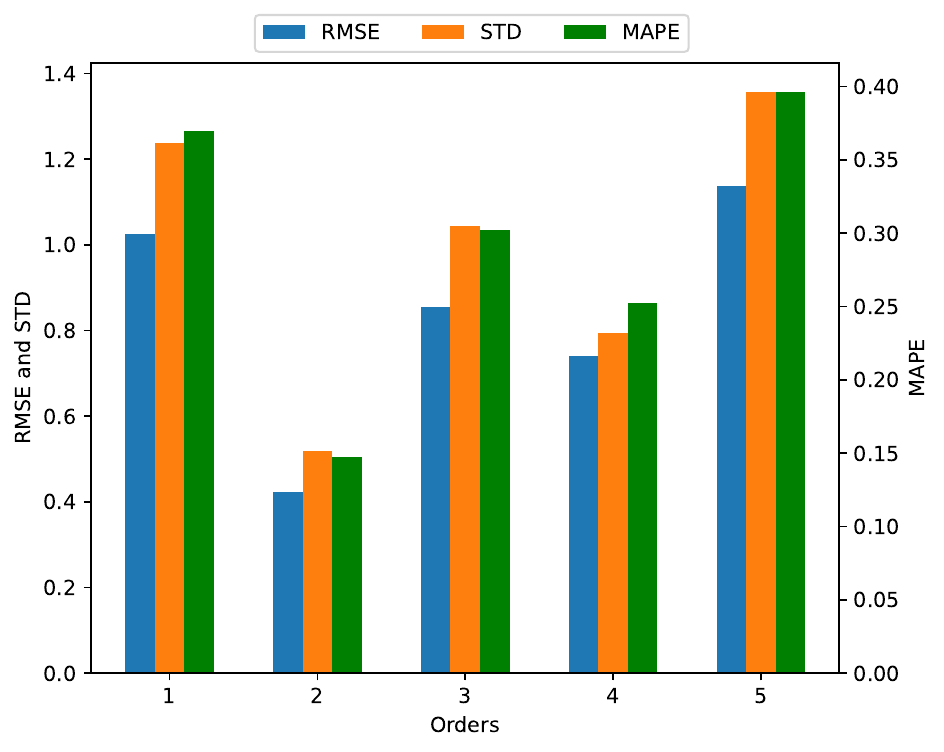}
    \caption{Different order of feature augmentation for CNC Turning dataset}
    \label{fig:order-CNC}
\end{figure}

\begin{figure}[!p]
    \centering
    \includegraphics[width=.35\textwidth]{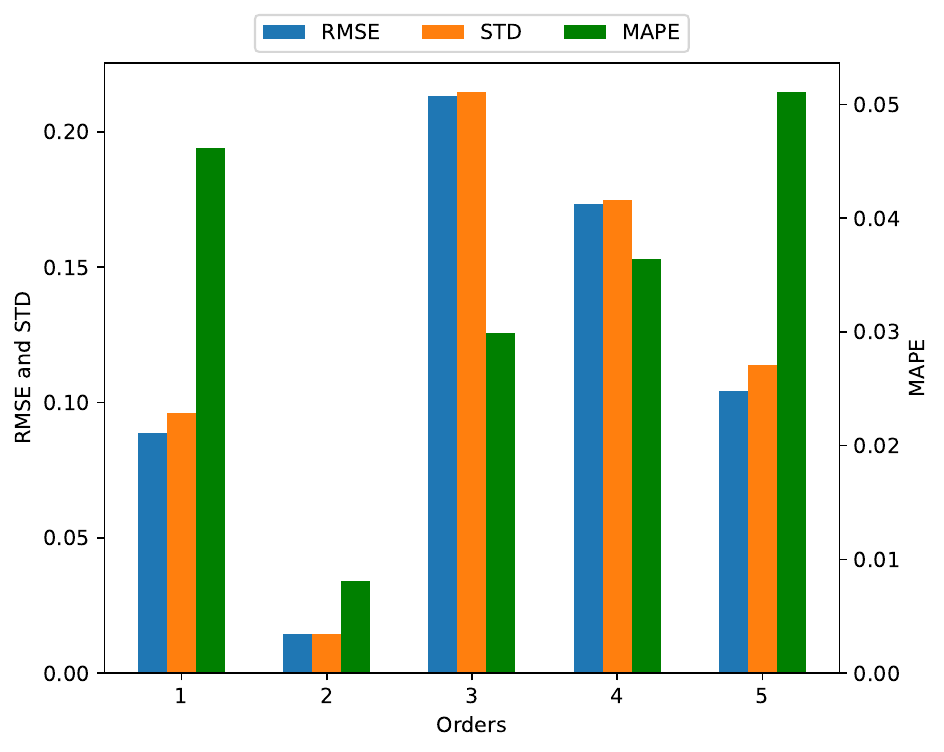}
    \caption{Different order of feature augmentation for Cutting Vibration dataset}
    \label{fig:order-Cut}
\end{figure}

Figures \ref{fig:order-MJP}, \ref{fig:order-CNC}, and \ref{fig:order-Cut} depict the algorithm's performance across three distinct datasets under varying orders of polynomial feature augmentation. Our empirical analysis uncovers that second-order feature augmentation is the most advantageous. This selection aligns with the notion that an appropriate augmentation order can facilitate a more accurate extraction of underlying relationships by the neural network. Simultaneously, an overemphasis on higher orders can potentially contribute to noise in the data, consequently diminishing the quality of the learned representations.

\begin{figure}[!p]
    \centering
    \includegraphics[width=.35\textwidth]{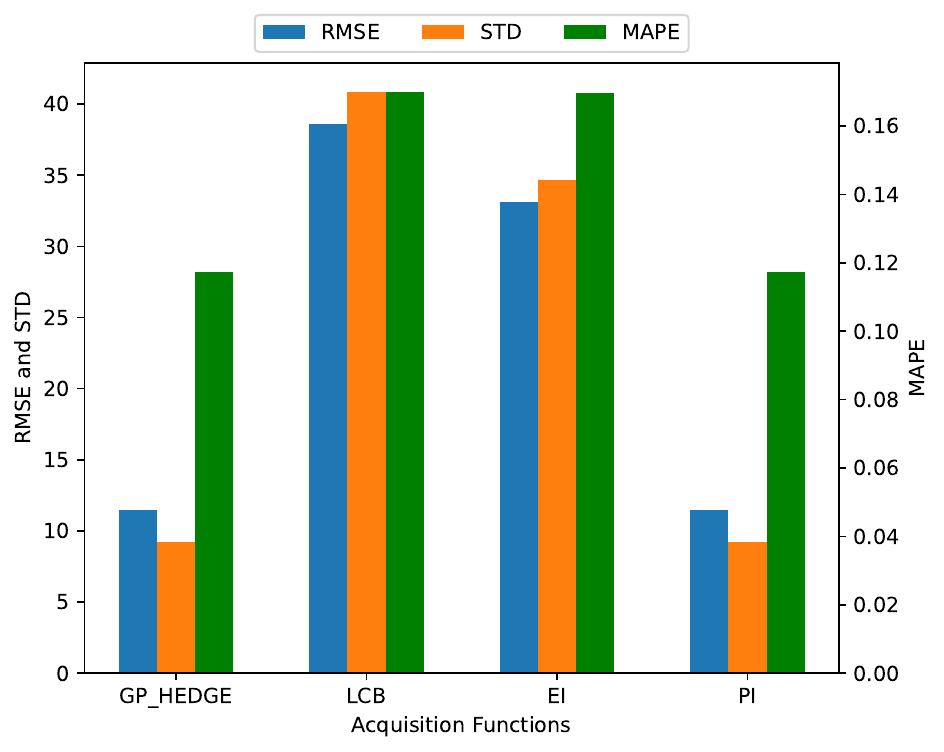}
    \caption{Different acquisition function for MJP dataset}
    \label{fig:acq-MJP}
\end{figure}

\begin{figure}[!p]
    \centering
    \includegraphics[width=.35\textwidth]{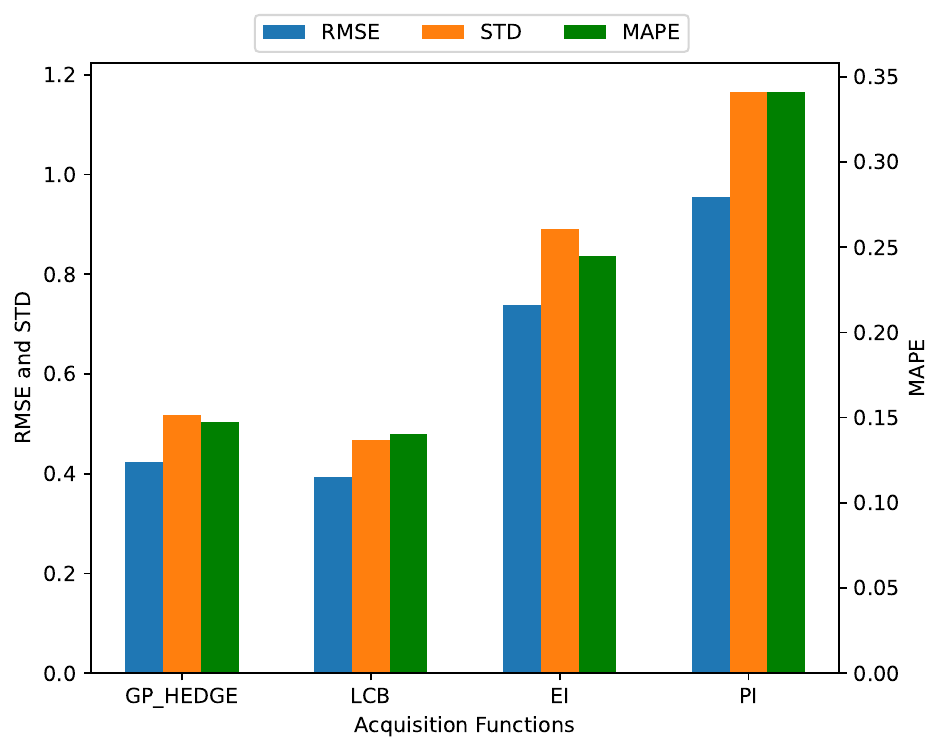}
    \caption{Different acquisition function for CNC Turning dataset}
    \label{fig:acq-CNC}
\end{figure}

\begin{figure}[!p]
    \centering
    \includegraphics[width=.35\textwidth]{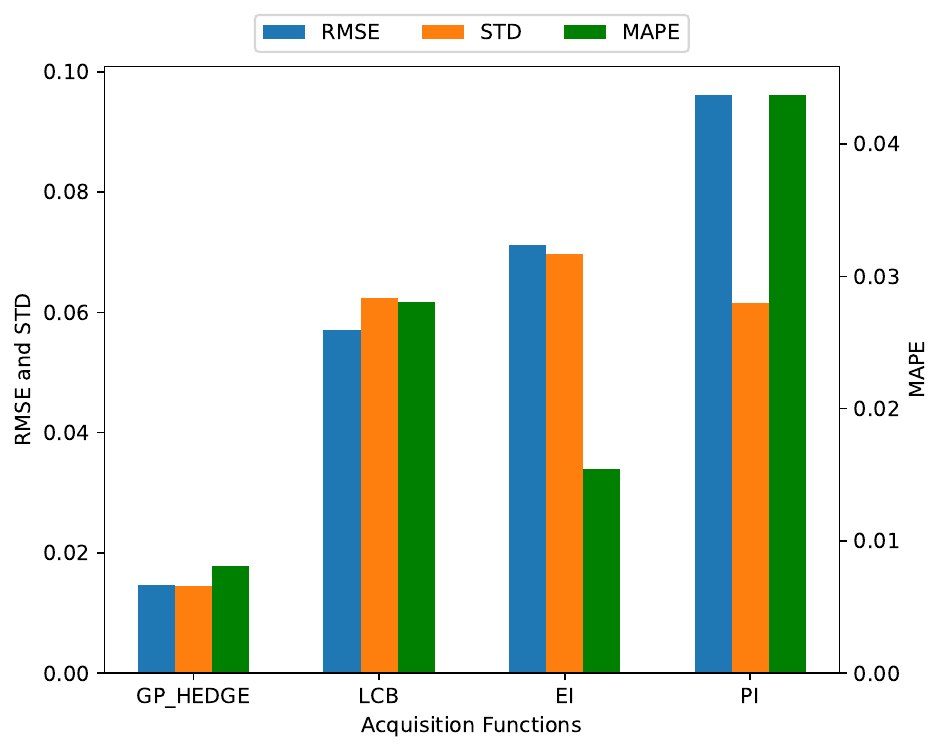}
    \caption{Different acquisition function for Cutting Vibration dataset}
    \label{fig:acq-Cut}
\end{figure}

Figures \ref{fig:acq-MJP}, \ref{fig:acq-CNC}, and \ref{fig:acq-Cut} showcase the performance metrics of various acquisition functions, namely GP\_HEDGE, LCB, EI, and PI, in different scenarios. Our evaluation reveals that the hybrid GP\_HEDGE outperforms other functions on the MJP and Cutting Vibration datasets. This superior performance can be ascribed to its ability to strike an effective balance between exploration and exploitation, thereby enabling efficient traversal of the search space within the Bayesian optimization process. However, in the context of the CNC Turning dataset, LCB demonstrates superior performance. This could be attributed to the complexity and presence of multiple optima within the landscape of the CNC Turning process, where LCB's tendency to explore regions with high uncertainty provides a performance advantage. These findings underscore the critical role of thoughtful acquisition function selection in fully leveraging the power of NAS.




\subsection{Ablation Study}
In this section, we conduct an extensive ablation study to examine the impact of the third stage within NASPrecision framework. As demonstrated in Table \ref{table:ablation}, there is a significant performance boost with the refinement stage, indicating that this stage effectively captures the details ignored by the previous stage.


\subsection{Discussion and Limitations}
While our proposed NASPrecision framework has shown remarkable accuracy and robustness in predicting surface roughness in ultra-precision machining, there are certain limitations that should be acknowledged.

\textbf{First}, our use of generative data augmentation has undoubtedly enriched the dataset and improved the robustness of the model, particularly for data-hungry neural networks. However, one should be aware that the generative model might synthesize data that does not correspond to real-world experimental conditions. This could introduce significant bias into the model, potentially affecting the generalization and applicability of the predictions. \textbf{Second}, the effectiveness of NAS is heavily dependent on the design of the search space. Ideally, a well-designed search space should encompass a wide range of functions, thereby exhibiting a high level of expressiveness. However, defining such a search space is a challenging task that requires careful consideration of the balance between generality and specificity. \textbf{Third}, it is worth noting that while NAS can deliver improved performance, it comes with a considerable computational cost. This renders our methods relatively time-consuming, especially when navigating larger search spaces. This issue is particularly pertinent in scenarios where rapid prediction is crucial or computational resources are limited.

\section{Conclusion}
This paper presents a novel and generic NAS-driven multi-stage learning framework, NASPrecision designed for the prediction of surface roughness in ultra-precision machining processes. In pursuit of a framework that is universally applicable across diverse machining processes, varying machinery, and different data distributions, we proposed to incorporate NAS to automatically discover the optimal architectures tailored to the specific task at hand. We then proceed a multi-stage training, starting with an initial training stage. Subsequently, we employ a refinement stage to further improve the performance and rectify the potential bias introduced. Furthermore, a salient issue in ultra-precision machining, namely the scarcity and imbalance of data, is effectively addressed through the use of generative data augmentation techniques. Evaluation of our proposed model was performed using three distinct datasets, providing a broad-based perspective of its performance and flexibility. The results convincingly demonstrate that our methods significantly outperform traditional machine learning algorithms as well as ensemble learning algorithms. To corroborate the significance of each component within our model, we also conducted an ablation study. The findings of this analysis provide crucial insights into the roles and impacts of the individual components of NASPrecision framework.

For our future work, we anticipate the incorporation of advanced techniques in NAS into our existing framework. Our intent is not only to enrich the search space with more diverse and potent architectures but also to enhance the efficiency of the search procedure. By implementing cutting-edge NAS strategies, we expect to evolve our model's performance through more granular optimizations, thereby improving the predictive accuracy even further.\\


%

\noindent\textbf{Declaration of Competing Interest}

The authors declare that they have no known competing financial interests or personal relationships that could have appeared to influence the work reported in this paper.


%






\bibliographystyle{apalike} 

\end{document}